\documentclass{IEEEcsmag}

\usepackage[colorlinks,urlcolor=blue,linkcolor=blue,citecolor=blue]{hyperref}
\expandafter\def\expandafter\UrlBreaks\expandafter{\UrlBreaks\do\/\do\*\do\-\do\~\do\'\do\"\do\-}
\usepackage{upmath,color}
\usepackage{todonotes}

\usepackage{listings}

\definecolor{codegreen}{rgb}{0,0.6,0}
\definecolor{codegray}{rgb}{0.5,0.5,0.5}
\definecolor{codepurple}{rgb}{0.58,0,0.82}
\definecolor{backcolour}{rgb}{0.95,0.95,0.92}

\let\othelstnumber=\thelstnumber
\def\createlinenumber#1#2{
    \edef\thelstnumber{%
        \unexpanded{%
            \ifnum#1=\value{lstnumber}\relax
              #2%
            \else}%
        \expandafter\unexpanded\expandafter{\thelstnumber\othelstnumber\fi}%
    }
    \ifx\othelstnumber=\relax\else
      \let\othelstnumber\relax
    \fi
}

\lstdefinestyle{customc}{
  belowcaptionskip=1\baselineskip,
  breaklines=true,
  frame=single,
  xleftmargin=0.35cm,
  xrightmargin=0.15cm,
  numbers=left,
  numbersep=5pt,  
  language=C,
  showstringspaces=false,
  basicstyle=\footnotesize\ttfamily,
  keywordstyle=\bfseries\color{green!40!black},
  commentstyle=\itshape\color{purple!40!black},
  identifierstyle=\color{blue},
  stringstyle=\color{orange},
}

\lstdefinestyle{customcArianeExploit1}{
  breaklines=true,
  frame=single,
  xleftmargin=0.4cm,
  xrightmargin=0.2cm,
  numbers=left,
  numbersep=5pt,  
  language=C,
  showstringspaces=false,
  basicstyle=\footnotesize\ttfamily,
  keywordstyle=\bfseries\color{green!40!black},
  commentstyle=\itshape\color{purple!60!black},
  identifierstyle=\color{blue},
  stringstyle=\color{yellow!50!black},
  morekeywords={asm},
  keywordstyle=[2]\bfseries\color{brown!60!black},
}

\lstdefinestyle{customcArianeExploit}{
  breaklines=true,
  frame=single,
  xleftmargin=0.4cm,
  xrightmargin=0.2cm,
  numbers=left,
  numbersep=5pt,  
  language=C,
  showstringspaces=false,
  basicstyle=\footnotesize\ttfamily,
  keywordstyle=\bfseries\color{blue},
  commentstyle=\itshape\color{green!50!black},
  identifierstyle=\color{black},
  stringstyle=\color{brown},
  morekeywords={asm},
  keywordstyle=[2]\bfseries\color{black},
}

\lstdefinestyle{customlog}{
  breaklines=true,
  frame=single,
  xleftmargin=0.35cm,
  xrightmargin=0.15cm,
  numbers=left,
  numbersep=5pt,  
  language=C,
  showstringspaces=false,
  basicstyle=\footnotesize\ttfamily,
  keywordstyle=\color{blue},
  commentstyle=\itshape\color{purple!40!black},
  identifierstyle=\color{blue},
  stringstyle=\color{orange},
  keywords=[2]{INFO},
  keywords=[3]{ERROR},x
  keywordstyle=[2]\bfseries\color{green!40!black},
  keywordstyle=[3]\bfseries\color{red!500!black},
}

\definecolor{verilogcommentcolor}{RGB}{0,124,0}
\definecolor{verilogkeywordcolor}{RGB}{49,49,255}
\definecolor{backcolor}{RGB}{237,237,237}
\definecolor{verilogsystemcolor}{RGB}{128,0,255}
\definecolor{verilognumbercolor}{RGB}{255,143,102}
\definecolor{verilogstringcolor}{RGB}{160,160,160}
\definecolor{verilogdefinecolor}{RGB}{128,64,0}
\definecolor{verilogoperatorcolor}{RGB}{0,0,128}
\definecolor{pointcolor}{RGB}{192,0,0} 
\lstdefinestyle{prettyverilog}{
   backgroundcolor=\color{backcolor},
   language           = Verilog,
   commentstyle       = \color{verilogcommentcolor},
   alsoletter         = \$'0123456789\`,
   literate           = *{+}{{\verilogColorOperator{+}}}{1}%
                         {-}{{\verilogColorOperator{-}}}{1}%
                         {@}{{\verilogColorOperator{@}}}{1}%
                         {;}{{\verilogColorOperator{;}}}{1}%
                         {*}{{\verilogColorOperator{*}}}{1}%
                         {?}{{\verilogColorOperator{? }}}{1}%
                         {:}{{\verilogColorOperator{:}}}{1}%
                         {<}{{\verilogColorOperator{<}}}{1}%
                         {>}{{\verilogColorOperator{>}}}{1}%
                         {!}{{\verilogColorOperator{!}}}{1}%
                         {^}{{\verilogColorOperator{^}}}{1}%
                         {|}{{\verilogColorOperator{| }}}{1}%
                         {||}{{\verilogColorOperator{|| }}}{1}%
                         {=}{{\verilogColorOperator{= }}}{1}%
                         {==}{{\verilogColorOperator{== }}}{1}%
                         {=>}{{\verilogColorOperator{=> }}}{1}%
                         {[}{{\verilogColorOperator{[}}}{1}%
                         {]}{{\verilogColorOperator{]}}}{1}%
                         {(}{{\verilogColorOperator{(}}}{1}%
                         {)}{{\verilogColorOperator{)}}}{1}%
                         {rightbracket}{{\verilogColorOperator{)}}}{1}%
                         {,}{{\verilogColorOperator{,}}}{1}%
                         {.}{{\verilogColorOperator{.}}}{1}%
                         {~}{{\verilogColorOperator{$\sim$}}}{1}%
                         {\%}{{\verilogColorOperator{\%}}}{1}%
                         {\&}{{\verilogColorOperator{\& }}}{1}%
                         {\&\&}{{\verilogColorOperator{\&\& }}}{1}%
                         {\#}{{\verilogColorOperator{\#}}}{1}%
                         {\ /\ }{{\verilogColorOperator{\ /\ }}}{3}%
                         {\ _}{\ \_}{2}%
                        ,
   morestring         = [s][\color{verilogstringcolor}]{"}{"},%
   identifierstyle    = \color{black},
   vlogdefinestyle    = \color{verilogdefinecolor},
   vlogconstantstyle  = \color{verilognumbercolor},
   vlogsystemstyle    = \color{verilogsystemcolor},
   basicstyle         = \small\fontencoding{T1}\ttfamily,
  columns=fullflexible, 
   keywordstyle       = \bfseries\color{verilogkeywordcolor},
   morekeywords      = {val, when, port, coverage, unique},
   numbers            = left,
   numbersep          = 5pt,
   tabsize            = 2,
   escapeinside       = {/*!}{!*/},
   upquote            = true,
   sensitive          = true,
   showstringspaces   = false, 
   frame              = single, 
   breaklines         = true,
   abovecaptionskip   = 0pt,
   belowcaptionskip   = 0pt, 
   xleftmargin        =0.35cm,
   xrightmargin       =0.15cm,
   captionpos         = b,
   emph               = {Point, Point0, Point1, Point2, Point3, Point4, Point5, Point6, Point7, Point8, Point9},
   emphstyle          =\color{pointcolor},
   emph               = {[2] STVEC,SCOUNTEREN,MSTATUS,MTVEC,ML1_ICACHE_MISS,ML1_DCACHE_MISS,MITLB_MISS,MDTLB_MISS,
                             MLOAD,MSTORE,MEXCEPTION,MEXCEPTION_RET,MBRANCH_JUMP,MCALL,MRET,MMIS_PREDICT,MSB_FULL,
                             MIF_EMPTY,MHPM_COUNTER_17,MHPM_COUNTER_18,MHPM_COUNTER_19,MHPM_COUNTER_20,MHPM_COUNTER_21,
                             MHPM_COUNTER_22,MHPM_COUNTER_23,MHPM_COUNTER_24,MHPM_COUNTER_25,MHPM_COUNTER_26,MHPM_COUNTER_27,
                             MHPM_COUNTER_28,MHPM_COUNTER_29,MHPM_COUNTER_30,MHPM_COUNTER_31}, 
   emphstyle          = {[2]\bfseries\color{verilogkeywordcolor}}
}

\makeatletter

\newcommand\language@verilog{Verilog}
\expandafter\lst@NormedDef\expandafter\languageNormedDefd@verilog%
  \expandafter{\language@verilog}
  
\lst@SaveOutputDef{`'}\quotesngl@verilog
\lst@SaveOutputDef{``}\backtick@verilog
\lst@SaveOutputDef{`\$}\dollar@verilog

\newcommand\getfirstchar@verilog{}
\newcommand\getfirstchar@@verilog{}
\newcommand\firstchar@verilog{}
\def\getfirstchar@verilog#1{\getfirstchar@@verilog#1\relax}
\def\getfirstchar@@verilog#1#2\relax{\def\firstchar@verilog{#1}}

\newcommand\addedToOutput@verilog{}
\lst@AddToHook{Output}{\addedToOutput@verilog}

\newcommand\constantstyle@verilog{}
\lst@Key{vlogconstantstyle}\relax%
   {\def\constantstyle@verilog{#1}}

\newcommand\definestyle@verilog{}
\lst@Key{vlogdefinestyle}\relax%
   {\def\definestyle@verilog{#1}}

\newcommand\systemstyle@verilog{}
\lst@Key{vlogsystemstyle}\relax%
   {\def\systemstyle@verilog{#1}}

\newcount\currentchar@verilog
  
\newcommand\@ddedToOutput@verilog
{%
   \ifnum\lst@mode=\lst@Pmode%
      \expandafter\getfirstchar@verilog\expandafter{\the\lst@token}%
      \expandafter\ifx\firstchar@verilog\backtick@verilog
         \let\lst@thestyle\definestyle@verilog%
      \else
         \expandafter\ifx\firstchar@verilog\dollar@verilog
            \let\lst@thestyle\systemstyle@verilog%
         \else
            \expandafter\ifx\firstchar@verilog\quotesngl@verilog
               \let\lst@thestyle\constantstyle@verilog%
            \else
               \currentchar@verilog=48
               \loop
                  \expandafter\ifnum%
                  \expandafter`\firstchar@verilog=\currentchar@verilog%
                     \let\lst@thestyle\constantstyle@verilog%
                     \let\iterate\relax%
                  \fi
                  \advance\currentchar@verilog by \@ne%
                  \unless\ifnum\currentchar@verilog>57%
               \repeat%
            \fi
         \fi
      \fi
   \fi
}

\lst@AddToHook{PreInit}{%
  \ifx\lst@language\languageNormedDefd@verilog%
    \let\addedToOutput@verilog\@ddedToOutput@verilog%
  \fi
}

\newcommand{\verilogColorOperator}[1]
{%
  \ifnum\lst@mode=\lst@Pmode\relax%
   {\bfseries\textcolor{verilogoperatorcolor}{#1}}%
  \else
    #1%
  \fi
}

\makeatother

\lstdefinestyle{mystyle}{
    commentstyle=\textit,
    keywordstyle=\textbf,
    stringstyle=\color{codepurple},
    basicstyle=\ttfamily,
    breakatwhitespace=false,         
    breaklines=true,      
    frame=single, 
    framexleftmargin=\parindent,
    captionpos=b,                    
    keepspaces=true,                 
    numbers=left,    
    numberstyle=\normalsize,
    stepnumber=1,
    numbersep=5pt,   
    xleftmargin=1.5\parindent,
    showspaces=false,                
    showstringspaces=false,
    showtabs=false,                  
    tabsize=2
}

\newcommand{\chatfuzz}{\textit{ChatFuzz}}
\newcommand{\hypfuzz}{\textit{HyPFuzz}}
\newcommand{\thehuzz}{\textit{TheHuzz}}
\newcommand{\psofuzz}{\textit{PSOFuzz}}
\newcommand{\mabfuzz}{\textit{MABFuzz}}
\newcommand{\rfuzz}{\textit{RFUZZ}}
\newcommand{\difuzz}{\textit{DIFUZZRTL}}

\newcommand{\morfuzz}{\textit{MorFuzz}}
\newcommand{\socfuzzer}{\textit{SoCFuzzer}}
\newcommand{\cascade}{\textit{Cascade}}

\newcommand{\boom}{{\tt BOOM}}
\newcommand{\cva}{{\tt CVA6}}

\newcommand{\systemverilog}{\textit{SystemVerilog}}
\newcommand{\verilog}{\textit{Verilog}}

\newcommand{\vcs}{\textit{VCS}}
\newcommand{\verilator}{\textit{Verilator}}
\newcommand{\modelsim}{\textit{Modelsim}}

\newcommand{\zenbleed}{{\textit{Zenbleed}}}
\newcommand{\reptar}{{\textit{Reptar}}}
\newcommand{\downfall}{{\textit{Downfall}}}

\newcommand{\rowhammer}{{\textit{Rowhammer}}}
\newcommand{\zombieload}{{\textit{Zombieload}}}
\newcommand{\foreshadow}{{\textit{Foreshadow-NG}}}

\usepackage{tikz}
\usepackage{xcolor}
\newcommand*\cc[1]{\tikz[baseline=(char.base)]{
            \node[shape=circle,fill,inner sep=2pt] (char) {\textcolor{white}{#1}};}}

\jvol{XX}
\jnum{XX}
\paper{8}
\doiinfo{10.1109/MSEC.2024.3365070}
\jmonth{Mar/Apr}
\jname{IEEE Security \& Privacy}
\jtitle{Special Issue on Memory Safety}
\pubyear{2024}

\begin{document}

\sptitle{Theme: Memory Safety}

\title{Fuzzerfly Effect: Hardware Fuzzing for Memory Safety}

\author{Mohamadreza Rostami}
\affil{Technical University of Darmstadt, Darmstadt, 64289, Germany, \\ mohamadreza.rostami@trust.tu-darmstadt.de}

\author{Chen Chen}
\affil{Texas A\&M University, Texas, 77843, USA, \\ chenc@tamu.edu}

\author{Rahul Kande}
\affil{Texas A\&M University, Texas, 77843, USA, \\ rahulkande@tamu.edu}

\author{Huimin Li}
\affil{Delft University of Technology, Delft, 2628XE, Netherlands, \\ H.Li-7@tudelft.nl}

\author{Jeyavijayan Rajendran}
\affil{Texas A\&M University, Texas, 77843, USA, \\ jv.rajendran@tamu.edu}

\author{Ahmad-Reza Sadeghi}
\affil{Technical University of Darmstadt, Darmstadt, 64289, Germany, \\ ahmad.sadeghi@trust.tu-darmstadt.de}
\markboth{THEME}{THEME}

\begin{abstract}\looseness-1Hardware-level memory vulnerabilities severely threaten computing systems. However, hardware patching is typically inefficient or highly difficult post-fabrication. This paper investigates the effectiveness of hardware fuzzing in detecting hardware memory vulnerabilities and highlights challenges and potential future research directions to enhance hardware fuzzing for memory safety.

\hfill \break
\end{abstract}

\maketitle

\chapteri{M}emory safety is foundational to the overall security of computing systems and becomes a significant risk when compromised, particularly impacting the integrity of critical systems.
The increasing complexity of software and hardware systems and the discovery of novel and severe cross-layer attacks based on software-exploitable hardware vulnerabilities have highlighted the multifaceted nature of memory safety. 
Cross-layer attacks bring to light the limitations inherent in software-based memory safety approaches. They underscore that safeguarding memory goes beyond countering traditional software memory corruption attacks, such as user-after-free, out-of-bounds access, and heap/stack overflow, which typically stem from operating system flaws or software implementations.

The recent hardware vulnerabilities, such as \zenbleed\footnote{Zenbleed. \url{https://lock.cmpxchg8b.com/zenbleed.html}} and \downfall\footnote{Downfall. \url{https://downfall.page}}, challenge the widely held belief that a secure system can be achieved exclusively through secure software. These vulnerabilities illustrate that attackers can exploit vulnerabilities in the underlying hardware even if the software is carefully developed to protect sensitive data. These attacks exploit inherent microarchitectural flaws in modern processors to comprise sensitive information even within secure software or in the presence of software-based defenses. Moreover, vulnerabilities like \zombieload{} and \foreshadow{} demonstrate that even cutting-edge hardware architectures with advanced protection mechanisms like Intel Software Guard Extension (SGX) can succumb to unforeseen attack vectors.$^{1-2}$

The growing need for high-performance hardware has led to the incorporation of advanced optimization features in processors, such as speculative execution and special buffers, and the development of system-on-chips (SoCs) with dedicated special-purpose processing units, e.g., artificial intelligence (AI) co-processor, crypto co-processor, and on-chip graphics processing unit (GPU). 
Integrating new complex features and components into processors and SoCs has significantly broadened the potential attack surface for adversaries, extending memory safety concerns beyond traditional consideration of Dynamic Random-Access Memory (DRAM) and cache since they encompass microarchitectural and architectural resources within the processor and SoC, including general-purpose registers, vector extension registers, control/status registers, and SoC memory-mapped buffers and registers.

Hence, there is a need for effective tools and methodologies to identify and mitigate such memory safety issues in hardware designs. 
Existing hardware mitigation approaches against memory vulnerabilities can be categorized into (i) mitigation approaches based on architectural vulnerabilities, such as IMIX, CURE, CHERI, Sanctum, and ARM Memory Tagging Extension (MTE). These methods provide isolation between processes in different levels of security and protect the system against most known memory-related vulnerabilities.
However, they are mostly limited to architectural leakages and overlook microarchitectural information leakages, including processors and co-processors internal buffers.$^{3-5}$ (ii) mitigation approaches based on microarchitectural vulnerabilities, such as CHUNKED-CACHE. While CHUNKED-CACHE can provide a secure cache by isolating different users' cache lines, it is focused on the specific microarchitectural memory component, i.e., cache,
and does not cover other memory components in the SoC.$^{2}$

Due to the limitations of mitigation approaches in preventing new vulnerabilities, which we will explain in detail in the following sections, it is crucial to have vulnerability detection techniques. However, unlike software vulnerabilities that can be digitally patched, hardware patches often require physical modifications or replacements. Consequently, it is crucial to detect these vulnerabilities proactively before the fabrication.$^1$
Existing hardware memory vulnerability detection approaches can be broadly divided into static and dynamic techniques. (a) Static techniques, such as formal verification, analyze hardware without requiring execution to identify potential vulnerabilities. However, these methods often face state-explosion problems, particularly for complex and large-scale hardware designs like processors and SoCs. (b) Dynamic techniques, such as hardware fuzzing, monitor hardware behavior during the execution of crafted testates to detect memory vulnerabilities.$^{6-15}$ %
Unlike static techniques, dynamic techniques are scalable to large designs, as we will explore in this paper.$^{6}$

Recently, hardware fuzzing has gained prominence as an effective dynamic vulnerability detection technique. It has emerged as a promising tool for identifying vulnerabilities in large-scale designs, especially those that may be overlooked by traditional detection and mitigation approaches, including issues related to memory.
An illustrative example of the efficacy of hardware fuzzing is the discovery of the \zenbleed{} memory vulnerability in AMD's Zen 2 microarchitecture, which was successfully detected using hardware fuzzing techniques. Moreover, a noteworthy observation is that memory-related vulnerabilities constitute a substantial portion of the vulnerabilities identified by state-of-the-art hardware fuzzers in open-source processors or SoCs.$^{6-15}$ For instance, 36\% of vulnerabilities detected by the hardware fuzzer \thehuzz{} rooted in hardware-level memory.$^{6}$ 
This highlights the potential of hardware fuzzing in uncovering memory-related vulnerabilities that other mitigation and detection approaches might overlook.

In this paper, we focus on hardware-related memory-safety issues and investigate the efficacy of innovative techniques, such as hardware fuzzing, in preemptively identifying vulnerabilities pre-fabrication. We will systematically review established hardware fuzzing techniques, explain how they overcome the limitations of existing memory-related vulnerability detection methodologies, and investigate their effectiveness in identifying memory-related vulnerabilities.
We compare various hardware fuzzing techniques, highlighting their unique strengths and weaknesses in detecting hardware-level memory vulnerabilities.
Afterward, we analyze and explain why hardware fuzzing has the potential to detect particular memory vulnerabilities overlooked by other methods. In conclusion, this paper will present potential directions for the future development of hardware fuzzers to enhance their capabilities in memory safety verification.

\section{Background}
\subsection{Memory safety in Software}
Memory safety vulnerabilities in software are inherent flaws that emerge within the software itself or the software platform that runs it, such as the operating system (OS) or firmware. Many vulnerabilities commonly originate from unsafe programming languages such as C and C++, while favored for system-level programming due to their efficiency and direct memory access. However, these languages have security-critical shortcomings, such as weak typing and low-level memory management. Consequentially, memory errors include spatial and temporal errors, often resulting in memory corruption attacks such as use-after-free, out-of-bounds access, and heap/stack overflow. Exploiting these vulnerabilities allows unauthorized access to sensitive information and jeopardizes the integrity of computing systems.

Therefore, addressing memory vulnerabilities is crucial to fortifying software security. Several tools have been developed to identify and rectify memory errors across the software development lifecycle. Static analysis tools scrutinize source code before execution, aiding in early vulnerability detection; however, their comprehensiveness across diverse scenarios remains a significant challenge as code complexity increases. In contrast, dynamic analysis tools track memory access and incorporate code to detect errors during runtime. However, their implementation can impact application performance and memory usage, particularly in real-time or high-performance applications. For example, the utilization of \textit{AddressSanitizer} can decrease performance by $73\%$ and increase memory consumption by $3.4\times$, attributable to additional code for runtime analysis.

To address these challenges, recent research has investigated hardware-based approaches to enhance memory safety, such as Capability Hardware Enhanced RISC Instructions (CHERI), In-Process Memory Isolation EXtension (IMIX), ARM Pointer Authentication Code (PAC), and Intel Control-flow Enforcement Technology (CET). These technologies aim to isolate distinct memory regions or objects, preventing unauthorized access and minimizing the impact of various memory-related vulnerabilities such as buffer overflows, stack smashing, and heap-based vulnerabilities. Implementing security features at the hardware level often incurs less overhead than dynamic analysis tools. However, integrating these hardware-based solutions necessitates adaptation across existing software architectures and systems, posing significant implementation challenges. Achieving compatibility and seamless integration across diverse software platforms remains a pertinent concern in the adoption of these hardware-based memory safety solutions.$^{3,5}$

Besides, hardware-level defense mechanisms have been developed and implemented to mitigate software vulnerabilities, including Address Space Layout Randomization (ASLR), Data Execution Prevention (DEP), and Control Flow Integrity (CFI). These techniques collectively provide an additional layer of protection by leveraging hardware-based features to fortify software and raise the complexity level for potential attackers. Nevertheless, they often necessitate hardware support, making deployment and adoption challenging, particularly for legacy systems lacking the required hardware features.

\subsection{Memory Safety in Hardware}

Recent discoveries of various hardware vulnerabilities have challenged the traditional assumption that system security solely relies on secure software practices. Memory errors extend beyond software, encompassing hardware vulnerabilities that significantly influence overall system security. Analyzing memory safety through a hardware lens requires delving into various levels of abstraction and threat models. In this context, memory safety is broadly classified into two principal domains: Architectural Memory Safety and Microarchitectural Memory Safety.

\noindent\textbf{Architectural Memory Safety} encompasses bugs, vulnerabilities, and mitigation strategies that directly impact the processor's architectural state. These issues are directly observable and can be addressed through architectural modifications. Architectural memory safety issues often stem from flaws in the processor's instruction set architecture (ISA) or the memory management unit~(MMU). These issues can lead to memory access errors, such as out-of-bounds memory access or unauthorized memory access. Mitigation strategies for architectural memory safety issues typically involve ISA modifications, MMU enhancements, or compiler-based techniques.

One prominent example of an hardware memory vulnerability is \rowhammer{} which specifically affects DRAM. \rowhammer{} operates by repeatedly accessing specific rows of memory, inducing electrical interference that causes bit flips the values of adjacent memory cells. This can lead to unauthorized data access and system-level compromisesincluding tampering with cryptographic keys, escalation of privileges, and causing denial of service attacks. 
Different mitigation techniques are proposed to mitigate \rowhammer{}; among these, Error-Correcting Codes (ECCs) utilize additional data bits to detect and rectify bit flips caused by \rowhammer{}, thereby fortifying memory integrity and minimizing the impact of such vulnerabilities.

Another recent notable example is the \reptar{}\footnote{Reptar. \url{https://lock.cmpxchg8b.com/reptar.html}} vulnerability which was detected through fuzzing. Researchers found that adding redundant prefixes, i.e., \texttt{rex.r}, to an FSRM-optimized repeating move operation, i.e., \texttt{rep movsb},  will lead the processor to an unstable state; as a result, branches jump to unexpected locations. This abnormality may lead to system crashes, hangs, and information leaks, even from unprivileged guest virtual machines. Intel fixed this vulnerability by providing a microcode update.

\noindent\textbf{Microarchitectural Memory Safety} deals with bugs and vulnerabilities that are not directly visible at the architectural level and require side-channel techniques to reveal their impact. 
Microarchitectural memory safety issues arise from implementation and design flaws in the processor's microarchitecture.
These issues can manifest in various ways, such as cache side-channel attacks or speculative execution attacks. Mitigating microarchitectural memory safety issues often involves microarchitectural modifications, such as speculative execution hardening and cache isolation techniques.

Notable examples of instances that highlight critical microarchitectural memory safety issues are \downfall{} and \zenbleed{}, which are transient execution vulnerabilities that were released in 2023 affecting Intel and AMD processors, respectively.

\downfall{} leverages the \texttt{gather} instruction within Intel ISA, bypassing previously implemented hardware fixes and security updates for transient execution-based vulnerabilities. The \texttt{gather} instruction family, an optimization designed to access scattered data in memory addresses, employs a shared buffer within a physical core. However, this shared buffer becomes vulnerable, potentially exposing the content of prior or concurrent executions of gather instructions within the transient window. Exploiting this vulnerability could enable malicious actors to extract sensitive information. 

\zenbleed{} exploits the zeroing register optimization, i.e., the \texttt{vzeroupper} instruction in AMD processors. The \texttt{vzeroupper} instruction aids processors in mitigating redundant register dependencies. However, a victim process executing the \texttt{vzeroupper} instruction within a mispredicted speculative window leaves the register undefined. Subsequently, other processes may access the undefined register, which contains random data from the victim process. As all processes share the register file, attackers can read registers from other processes, potentially compromising sensitive information and threatening system security. Patches to these two bugs have been provided by Intel and AMD, respectively.

Architectural mitigation strategies from industry and academia, such as ARM MTE, AMD SEV, Intel's TDX, Santum, Keystone, and CURE, have played a crucial role in fortifying system security against known memory-related vulnerabilities.$^4$ These approaches primarily target architectural memory vulnerabilities and create isolation between processes of varying security levels. However, their effectiveness is limited as they do not consider microarchitectural vulnerabilities 
,leaving potential security gaps unaddressed. Microarchitectural mitigation methods, like CHUNKED-CACHE, are more targeted and focus on protecting specific memory parts like caches.
Nevertheless, these methods do not cover all microarchitectural memory elements within the hardware.$^2$

Given the complexity of providing patches for newly discovered hardware vulnerabilities, there has been a noticeable shift toward pre-silicon detection methodologies. Static vulnerability detection techniques, such as formal verification, gained attention initially.
However, these methods encounter significant challenges, such as state-explosion issues, especially in complex hardware structures like processors and SoCs. Consequently, dynamic vulnerability detection techniques, particularly hardware fuzzing, have emerged. Hardware fuzzing involves the execution of test cases to monitor hardware behavior and identify memory vulnerabilities actively. Unlike static methods, hardware fuzzing exhibits better scalability.
Moreover, hardware fuzzing has emerged as a promising tool in efficiently identifying vulnerabilities in extensive designs that traditional detection and mitigation approaches may overlook, particularly those associated with memory, for instance, the detection of \zenbleed{}.

In this paper, we aim to investigate the efficacy of hardware fuzzing in detecting and mitigating hardware memory vulnerabilities.

\section{Hardware Fuzzing for Memory Safety}

\begin{figure*}
    \centering
    \includegraphics[width=\linewidth]{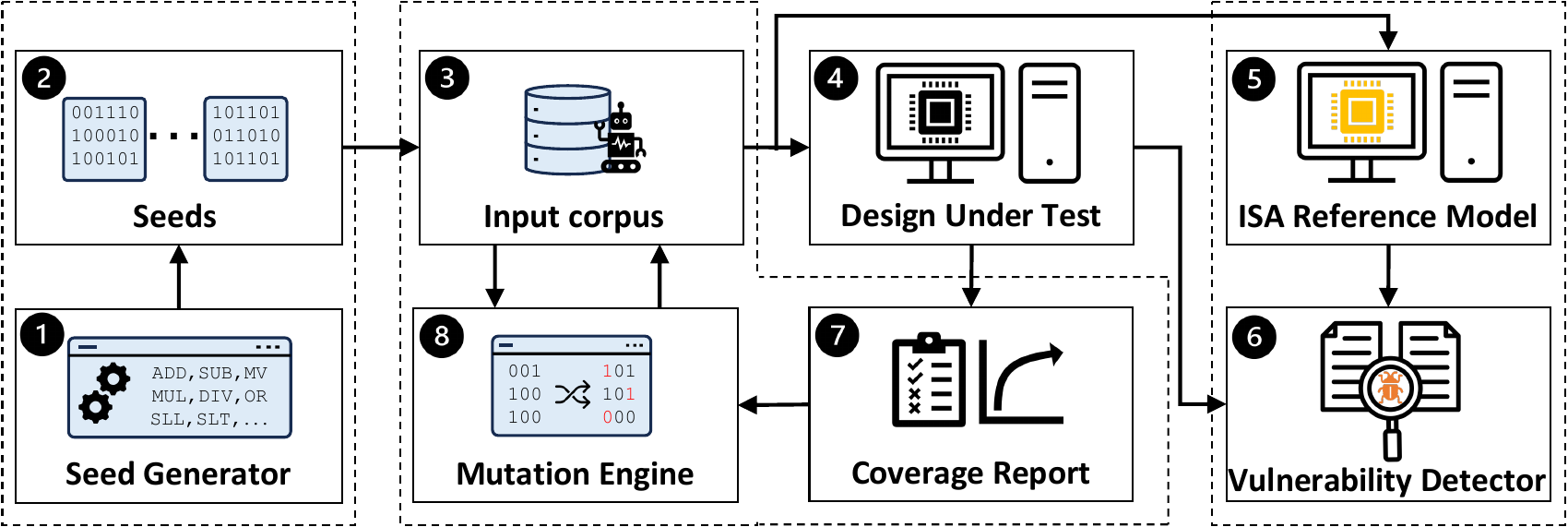}
    \caption{Hardware Fuzzing Framework.}
    \label{fig:fuzz_des}
\end{figure*}

Fuzzing is a dynamic technique widely used to discover coding errors and security holes. 
It involves providing invalid, unexpected, or random data as inputs to a design. 
The main goal is to observe the design's response and identify abnormal results that are potentially caused by vulnerabilities. 
Fuzzing operates under the premise that unexpected or edge-case inputs can lead to unanticipated behavior, exposing vulnerabilities that standard testing might overlook.
We highlight the advantages of using hardware fuzzers for memory safety next.$^{6-15}$

\subsection{Advantages of Hardware Fuzzers}
Hardware fuzzing offers several distinct advantages over other approaches:

\noindent\textbf{Automation.} 
Unlike static techniques, hardware fuzzing does not require manual code inspection or manually written security properties.
Hardware fuzzers automatically generate inputs, execute them on the design under test~(DUT), collect activity in the DUT, analyze the output of the DUT, and flag potential vulnerabilities in the DUT. 
This automation significantly reduces the time and effort required to identify memory safety issues.

\noindent\textbf{Compatibility.} 
Hardware fuzzers can fuzz the DUTs both in the presence and absence of DUT source codes. 
Hardware fuzzers using industrial verification tools can easily integrate into the existing industrial verification flow.
This compatibility enables the detection of memory safety vulnerabilities at both pre- and post-silicon phases with minimal false positives.

\noindent\textbf{Efficiency.} 
Hardware fuzzers generate unexpected inputs aiming to explore unanticipated behaviors of a design.
Such generation of inputs ensures fast and efficient exploring of a 
wide range of execution paths and design spaces. 
This comprehensive coverage increases the likelihood of uncovering hidden vulnerabilities.
Moreover, the inputs not only discover but also ensure the reproducibility of the vulnerabilities, which helps engineers identify the root cause of vulnerabilities.

\noindent\textbf{Scalability.} Hardware fuzzing can be adapted to larger, more complex, or diverse hardware systems than formal verification while maintaining or improving the efficiency and effectiveness of vulnerability detection. 
As hardware systems become more complex, the state space that needs to be verified exponentially increases. This makes hardware fuzzing more applicable for memory safety in modern systems. 

Though hardware fuzzing cannot guarantee the completeness of verification and the absence of vulnerabilities, its automation, compatibility, efficiency, and scalability make it a promising strategy for verifying memory safety in modern, complicated systems.

\subsection{Hardware Fuzzers Workflow}
We explain the general working flow of hardware fuzzers in this section.

Figure~\ref{fig:fuzz_des} shows the core components of a hardware fuzzer: a \textit{seed generator}, \textit{input corpus}, \textit{mutator}, \textit{coverage feedback}, \textit{DUT}, and \textit{vulnerability detector}.
The seed generator \cc{1} generates an initial set of inputs called \textit{seeds} \cc{2} either manually or randomly. 
The fuzzer runs the DUT \cc{4} with these inputs \cc{3}, collects \textit{coverage} \cc{7}, and \textit{mutates} all ``interesting'' inputs (i.e., inputs that achieve coverage) using its mutator \cc{8} to generate new inputs and store them in the input corpus \cc{3}. 
Mutators are data manipulation operations, such as \textit{bit-flip}, \textit{byte-flip}, \textit{clone}, and \textit{swap}.
The vulnerability detector \cc{6} reports any vulnerabilities detected during the execution. 
The fuzzer executes these new inputs and repeats the cycle until it achieves the desired coverage. 
Next, we explain the various components and tasks performed by hardware fuzzers.

\noindent\textbf{DUT} is a hardware design at either the pre- or post-silicon stage.
The former provides source codes written in hardware description languages~(HDL) like \verilog{} and \systemverilog{}, hardware construction languages~(HCLs) like Chisel, or high-level synthesis~(HLS) like C. 
Hardware fuzzers use simulation tools like \verilator{}, Synopsys \vcs{}, and Siemens \modelsim{} to simulate or use field programmable gate arrays (FPGA) to emulate DUTs. DUT at the post-silicon stage is a fabricated hardware chip. %

\noindent\textbf{Coverage feedback.} Based on the availability of source codes, hardware fuzzers can be classified as white-box and black-box fuzzers and measure various types of hardware behaviors as coverage feedback.
Coverage points are assigned to each of these behaviors.
For example, in white-box fuzzing, branch coverage indicates whether the different paths of a branch statement are covered or not. 
Whenever a design enters one of the branch paths, its corresponding coverage point is considered covered; otherwise, it remains uncovered. While, in black-box fuzzing, due to the lack of source codes, outputs of a DUT are used as coverage feedback. When fuzzing commercial processors, Google uses hardware performance counters (HPCs) as coverage feedback to detect the \zenbleed{} vulnerability.

\noindent\textbf{Vulnerability detection.} 
Unlike software, hardware does not have events like crashes, memory leaks, and buffer overflows to flag vulnerabilities.
Instead, vulnerability detection in hardware fuzzers involves either assertion checking or differential testing.$^{6-15}$

In assertion checking, verification engineers insert the conditions to trigger the vulnerabilities or assertion properties into the DUT based on its specification and use the violations of these assertions during the simulation to detect vulnerabilities. 
However, assertion checking requires access to source code and is only applicable to white-box fuzzing. Moreover, assertion checking requires a specification for memory safety and hence cannot detect unknown vulnerabilities.

In differential testing, hardware fuzzers compare the outputs of the DUT and a golden reference model~(GRM), tested with the same input, to detect vulnerabilities.$^{6-10,12-14}$
Differential testing can be applied at both pre-silicon and post-silicon stages. 
Also, since it does not require writing assertions, it can detect unknown vulnerabilities.
GRMs are designed at a higher abstraction level to avoid vulnerabilities and, hence, generate expected outputs;
any mismatches between a DUT's and a GRM's outputs reveal potential vulnerabilities in the DUT.
For example, when fuzzing a processor, hardware fuzzers compare the processor's architecture states (e.g., values of general-purpose registers, memory, etc.) with the corresponding states of an ISA simulator.
The mismatches between the processor's and simulator's outputs represent potential vulnerabilities in the processor.

\subsection{Existing Hardware Fuzzers}
Several hardware fuzzing techniques have been developed to promote architectural memory safety at the pre-silicon stages.
We categorize existing hardware fuzzers based on input generation, configuration, platform, and vulnerability detection, as shown in Table~\ref{tab:fuzz_sum}.

\noindent\\
\textbf{Input generation.} Hardware fuzzers use various seed generation and mutation mechanisms, such as randomly generating seeds and using pre-defined mutators to generate subsequent inputs. 
\thehuzz{}, \psofuzz{}, \mabfuzz{}, \rfuzz{}, \difuzz{}, and \socfuzzer{} use this strategy.$^{6,8,9,11,12,15}$
However, it is hard to trigger advanced optimization features in processors, such as speculative execution, by randomly generating instructions,
as precise instruction order is crucial for triggering these advanced functionalities. Also, there is a deficiency in data and control flow entanglement. 
For instance, the data loaded by a load instruction may never be used by subsequent instructions. 
These two limitations constrain the capability of these hardware fuzzers to achieve high coverage and discover complex vulnerabilities, especially those related to memory.

To address these limitations, fuzzers use user-defined seed generation and mutation templates.
The template divides an input into different testing blocks, aiming to explore various functionalities of the processor.
For example, \morfuzz{} develops templates based on the functions of instructions.$^{13}$
A seed will contain multiple testing blocks following different templates. 
\morfuzz{} then applies mutators specialized for each template to generate new inputs.
Similarly, \cascade{} steers the control flow of testing blocks to make an input execute longer. 
\cascade{} clusters instructions based on their functionalities, selects instructions, and constrains their operands' values so that the execution of the input will either terminate or continue.$^{14}$

However, creating user-defined templates manually requires experts knowledgeable about processor functions, a time-consuming and error-prone process. 
\chatfuzz{}, on the other hand, uses Large Language Models (LLMs) to understand machine language, encompassing machine codes, ISA instructions, and their potential combinations.
Subsequently, it generates complex test cases with entangled data and control flow. \chatfuzz{} achieves the same coverage as state-of-the-art hardware fuzzers but does so 30$\times$ faster. Furthermore, \chatfuzz{} demonstrates its ability to trigger more complex features, such as the order of memory exceptions and atomic memory operations.$^{10}$

Similar to stochastic processes in other fields, the above fuzzers often have a low probability of generating inputs for logic requiring specific conditions---namely, the hard-to-reach design spaces of a fuzzer.
Memory access usually requires hardware to be in a specific privileged state. 
Failing to verify this logic leaves opportunities for severe memory vulnerabilities. 
To tackle this limitation, \hypfuzz{} combines formal verification with fuzzing, establishing a hybrid hardware fuzzer.$^7$
Formal tools are used to generate seeds that explore hard-to-reach design spaces. 
These seeds then lead the fuzzing process to explore the hard-to-reach design spaces further. 

\noindent\\
\textbf{Configuration.} This category classifies fuzzers based on how they schedule the seed generation and mutation for the next iteration. 
Fuzzers with static configuration use constant probability distributions for selecting instructions (during seed generation) and mutators (during input mutation),
irrespective of the design space explored or vulnerabilities detected.
The static configuration, however, slows the coverage achievement and reduces the fuzzer's efficiency in detecting vulnerabilities.
To overcome this limitation, \psofuzz{} uses particle swarm optimization (PSO) to dynamically identify the optimal mutators and instruction opcodes for seed generation.$^{8}$
Similarly, \mabfuzz{} uses the multi-armed bandit (MAB) algorithms to select seeds with the potential to generate more ``interesting'' inputs.$^{9}$
\psofuzz{} and \mabfuzz{} detect existing vulnerabilities faster than the base-fuzzers they use with static configurations.

\noindent\\ 
\textbf{Platform.} 
Hardware fuzzers detect vulnerabilities by executing DUTs on different platforms. 
FPGA emulation facilitates faster execution than RTL simulation but cannot generate code coverage like RTL simulators.
\rfuzz{} introduces an FPGA framework for hardware fuzzing, featuring a coverage metric that monitors the toggling of the select signal of each \texttt{2:1} multiplexer.${11}$
However, the instrumentation overhead of \rfuzz{} limits its deployment on complex processors.  
\difuzz{} simplifies \rfuzz{}'s coverage metric by monitoring the toggling of control registers.
However, \difuzz{}'s coverage metric still introduces high instrumentation overhead as the size and number of control signals increase, limiting its scalability.
Moreover, the coverage metric monitors partial combinational logic in hardware, leaving spaces for vulnerabilities.$^{12}$
\socfuzzer{} introduces an FPGA framework to verify SoC designs.$^{15}$
However, it relies on the coverage of input and output spaces for feedback to guide input generation, which does not directly reflect design space exploration.

Fuzzers based on RTL simulators use intrinsic code coverage metrics as feedback.  
For example, \thehuzz{} leverages coverage metrics from commercial simulators, such as Synopsys \vcs{} to guide the fuzzer. 
These coverage metrics include \textit{branch}, \textit{finite-state machine (FSM)}, and \textit{toggle}, to monitor both combinational and sequential logic in hardware, guiding \thehuzz{} to explore design spaces and detect vulnerabilities exhaustively.$^6$
However, their simulations are not as fast as FPGA emulation.

\noindent\\
\textbf{Vulnerability detection.}
Existing hardware fuzzers apply either assertion checking or differential testing to detect vulnerabilities.
\rfuzz{} and \socfuzzer{} apply assertion checking and can only detect vulnerabilities that violate security specifications or known conditions. $^{11,15}$
\difuzz{} and \thehuzz{} deploy GRMs to detect previously unknown vulnerabilities on processors by differential testing.$^{6,11}$
However, the differential testing results contain many false positives due to the differences between the setup of GRMs and simulation environments, such as the memory allocation of stack and heap.
Also, due to the massive amount of inputs, identifying the specific sequences of instructions that cause vulnerabilities is difficult.
\morfuzz{} reduces the false positives of differential testing by synchronizing the setup of GRMs and simulation environments. \cascade{} identifies the minimal sequence of instructions that cause a vulnerability by systematically omitting instructions until the instruction sequence controlling the vulnerability behavior is detected.$^{13-14}$

\begin{table}[]
\caption{Summary of existing hardware fuzzers.}
\label{tab:fuzz_sum}
\resizebox{\columnwidth}{!}{%
\begin{tabular}{|c|c|c|c|c|}
\hline
\textbf{Method} & \textbf{\begin{tabular}[c]{@{}c@{}}Input\\ generation\end{tabular}} & \textbf{Configuration} & \textbf{Platform} & \textbf{\begin{tabular}[c]{@{}c@{}}Vulnerability\\ detection\end{tabular}} \\ \hline
\thehuzz{}$^6$         & Stochastic                                                              & Static            & Simulator         & \begin{tabular}[c]{@{}c@{}}Differential\\ testing\end{tabular}             \\ \hline
\hypfuzz{}$^7$         & Formal-assisted                                                     & Static            & Simulator         & \begin{tabular}[c]{@{}c@{}}Differential\\ testing\end{tabular}             \\ \hline
\psofuzz{}$^8$         & Stochastic                                                              & Dynamic           & Simulator         & \begin{tabular}[c]{@{}c@{}}Differential\\ testing\end{tabular}             \\ \hline
\mabfuzz{}$^9$         & Stochastic                                                              & Dynamic           & Simulator         & \begin{tabular}[c]{@{}c@{}}Differential\\ testing\end{tabular}             \\ \hline
\chatfuzz{}$^{10}$        & LLM-assisted                                                        & Static            & Simulator         & \begin{tabular}[c]{@{}c@{}}Differential\\ testing\end{tabular}             \\ \hline
\rfuzz{}$^{11}$           & Stochastic                                                              & Static            & FPGA              & \begin{tabular}[c]{@{}c@{}}Assertion\\ checking\end{tabular}               \\ \hline
\difuzz{}$^{12}$       & Stochastic                                                              & Static            & FPGA              & \begin{tabular}[c]{@{}c@{}}Differential\\ testing\end{tabular}               \\ \hline
\morfuzz{}$^{13}$         & Template                                                            & Static            & Simulator         & \begin{tabular}[c]{@{}c@{}}Differential\\ testing\end{tabular}             \\ \hline
\cascade{}$^{14}$         & Template                                                            & Static            & Simulator         & \begin{tabular}[c]{@{}c@{}}Differential\\ testing\end{tabular}             \\ \hline
\socfuzzer{}$^{15}$       & Stochastic                                                              & Static            & FPGA              & \begin{tabular}[c]{@{}c@{}}Assertion\\ checking\end{tabular}               \\ \hline
\end{tabular}%
}
\end{table}

\subsection{Hardware Memory Vulnerabilities Identified by Hardware Fuzzers}

Existing fuzzers detected numerous memory-related vulnerabilities by fuzzing popular open-source RISC-V processors. We now discuss some of these vulnerabilities and describe how these vulnerabilities in hardware can be exploited through software. 

\noindent\textbf{\textit{FENCE.I} instruction vulnerability.}
This is a vulnerability in the RISC-V ISA-based \cva{} processor, where the decoder of the processor fails to identify critical, memory management-related \textit{FENCE.I} instructions. 
This vulnerability is caused by the incorrect implementation of decoding logic for the \textit{FENCE.I} instruction and is similar to the expected behavior violation vulnerability, CWE-440. 
The implementation of the decoder module in the CVA6 processor includes additional and incorrect constraints when detecting the \textit{FENCE.I} instruction. %
As a result, the CVA6 processor fails to recognize some of the valid \textit{FENCE.I} instructions with non-zero values in its \textit{imm} and \textit{rs1} fields. 

The RISC-V ISA requires programs to use \textit{FENCE.I} instructions 
when performing memory-critical operations, such as updating instruction memory to ensure cache coherence in the processor. The hardware performs actions such as flushing the instruction pipeline and cache lines to achieve cache coherence. 
Thus, this vulnerability makes even a memory-safe software program (that uses \textit{FENCE.I} instructions) vulnerable to memory-safety attacks due to cache incoherence. 

Triggering this vulnerability requires verifying the processors with different operand values for the \textit{FENCE.I} instruction. 
Formal verification is impractical due to the vast number of instructions and operand values defined in the RISC-V ISA, making it difficult to verify all possible assertions.$^{6-7}$
In contrast, fuzzers are better suited for detecting such vulnerabilities as they employ mutation techniques to vary instruction operand values. 

\noindent\textbf{Cache coherency vulnerability.} 
Existing fuzzers detected another memory-related vulnerability related to the cache coherency in the CVA6 processor. 
This issue arises when the software program modifies instruction memory, and the processor fails to use updated instructions when executing from the modified memory location, contrary to RISC-V specification expectations.
The vulnerability stems from the CVA6 processor not having hardware mechanisms to detect or prevent memory incoherence, relying instead on software to use \textit{FENCE.I} instructions for memory safety operations.
While the ISA specification does not require hardware to ensure memory safety in this scenario, it is recommended that hardware implementation also include memory-safety checks for instances when either the software programs miss the \textit{FENCE.I} instruction or when the \textit{FENCE.I} instruction is implemented incorrectly in hardware (such as in the case of the vulnerability discussed above). 
Failing to have such memory-safety mechanisms in hardware can lead to the execution of incorrect instructions and cause memory and storage vulnerability, CWE-1202.$^{6-7,10}$

Triggering this vulnerability involves verifying the processor with memory-sensitive operations like overwriting instruction memory.
Fuzzers are effective in identifying these vulnerabilities due to their random instruction generation and mutation, capable of creating programs to modify instruction memory.

\noindent\textbf{Register value vulnerability.} 
This is a severe memory safety vulnerability in the \cva{} processor that returns unknown/random values when accessing the values of control and status registers (CSR). 
Such registers can be hardware performance counters~(HPCs). Designers widely use this type of register to identify the performance bottleneck of programs and detect malware.
It is a cross-module vulnerability that a module controlling the access of the CSRs assumes there are 32 registers. In comparison, another module storing the values of the CSRs only allocates 16 registers.
When a program accesses a CSR with no corresponding register, the processor returns unknown/random values.
This vulnerability\footnote{\url{https://nvd.nist.gov/vuln/detail/CVE-2022-33021}} resulted in \texttt{CVE-2022-3302} and can be categorized in CWE-1281 because the unknown/random value will cause unexpected memory behaviors in processors. 

This vulnerability is difficult to detect by non-hybrid fuzzing techniques because fuzzing alone has a low probability of generating inputs to access the correct CSR. At the same time, formal verification alone faces the state explosion issue from exploring all CSRs.
\hypfuzz{} detected this vulnerability using a formal tool to generate an input that accesses one CSR and then using the fuzzer to mutate the input to access the rest of the CSRs.$^7$

\section{Discussion and Future Direction}

Hardware fuzzing, while promising, presents substantial challenges and gaps in current methodologies that must be addressed.

Academic hardware fuzzers commonly utilize open-source hardware and processors as benchmarks. However, the absence of industry-level standards means that current open-source processors lack specific advanced functionalities like optimization buffers and trusted execution environments. 
Consequently,  open-source designs may not accurately represent real-world processors and SoCs.
Some hardware fuzzers, e.g.,  \cascade{},
are reporting the detection of functional bugs and security vulnerabilities in outdated and decommissioned open-source processor implementations, e.g., Kronos RISC-V and PicoRV32, rendering them irrelevant to industry standards. 
It's important to note that a key advantage of hardware fuzzers lies in their ability to assess advanced, complex, and real-world scale processors. Thus, utilizing simple and small designs as benchmarks to demonstrate hardware fuzzer effectiveness is understating the potential of hardware fuzzing. In cases involving small designs, alternative methodologies, such as formal methods, can offer superior assurances and coverage.

Indeed, we witnessed this gap between industry-standard full-fledged processors/SoCs and the open-source hardware in the world's largest SoC security competition, Hack@EVENT, a joint academia-industry collaboration since 2018\footnote{\url{https://hackthesilicon.com}}. The competition requires teams from both industry and academia to detect and analyze real-world hardware vulnerabilities that are deliberately injected into open-source SoC design. These vulnerabilities are emulated and inspired by the hardware Common Weakness Enumeration (CWEs) from MITRE corporation, which can serve as a benchmark for evaluating security tools and establish a foundational reference for identifying, mitigating, and preventing security vulnerabilities. Additionally, to enhance the representativeness of CWEs, we provide demonstrative examples for hardware CWEs in collaboration with the MITRE Corporation\footnote{So far our joint work has resulted in demonstrative examples for over 20\% of hardware CWEs which can be found in \url{https://cwe.mitre.org}}.

Hence, to develop a hardware fuzzer aligned with industry standards, researchers must rigorously evaluate their hardware fuzzers against the most complex open-source processors with active communities like \boom{} and try to identify bugs representing different types of CWEs. One good mature benchmark would be Hack@EVENT benchmarks, a comprehensive set of industry-standard vulnerabilities in hardware for system security research, to democratize the hardware security research for detecting vulnerabilities; without these testbeds, only elite companies who design real-world design can evaluate their security evaluation tools.

Through systematic analysis of existing hardware fuzzers and careful consideration of the capabilities inherent in hardware fuzzing, we have identified potential research directions to advance hardware fuzzing for memory safety.  

\subsection{Vulnerability Detection}
The efficacy of vulnerability detection within the hardware fuzzing framework relies on identifying discrepancies from comparing execution traces derived from simulating the hardware against those from the hardware's GRM. 

However, this vulnerability detection mechanism possesses the following limitations:

\noindent\textbf{Discrepancies vs. vulnerabilities.} Discrepancies do not necessarily indicate vulnerabilities. Hence, certain deviations from expected behavior, as specified by GRM, may arise due to the flexibility defined in the ISA specification.$^{10}$ These mismatches increase false positives within the detection mechanism, thereby augmenting the manual workload associated with vulnerability detection.

\noindent\textbf{Golden reference models vs. microarchitectural behaviors.} GRMs fall short in accurately capturing the design-under-test, the primary root cause for recent critical vulnerabilities, particularly those associated with memory. This limitation underscores the necessity to explore microarchitectural behaviors for vulnerability detection, offering the potential for hardware fuzzers to identify more severe vulnerabilities, including those related to advanced features such as transient execution and buffer optimizations. A plausible approach to addressing this limitation involves integrating information flow tracking techniques alongside hardware fuzzing to identify microarchitectural memory vulnerabilities.$^{1-2,6-14}$

\noindent\textbf{Golden model vs. no golden model.} The absence of a GRM is a recurrent challenge, particularly evident in the open-source community, where an SoC with identical peripherals may lack a corresponding golden model.
Consequently, developing a fuzzer that operates independently of a GRM seems imperative.

\subsection{Pinpointing the Vulnerability}
Current hardware fuzzing frameworks cannot identify the root cause of the vulnerability within the design, i.e., its location in the hardware design. The expansive code space in large hardware designs, such as processors and SoCs, poses a significant challenge in pinpointing the source of a vulnerability. This task requires manual intervention and an in-depth comprehension of the microarchitectures underlying the design. 
An idea to explore is the integration of Information Flow Tracking (IFT) techniques into hardware. IFT is a methodology that observes the flow of information within hardware
registers and modules.
So, upon detecting a vulnerability during the fuzzing process, the hardware fuzzer can utilize the information provided by the IFT process to trace back the leaked information back to its root cause.

\subsection{Hardware Fuzzing for SoC}
In contrast to individual processors or standalone hardware designs, there is a compelling need to enhance hardware fuzzers to validate SoC designs. 
This necessity arises because SoCs are more complicated than processors, which may have many vulnerabilities.

The peripheral components within SoCs encompass diverse memory modules, including Read-Only Memory (ROM), primary memory, numerous internal buffers, and secure FUSE memory designed to resist tampering. These memory modules store sensitive information, such as boot sequence instructions, private user data, and cryptographic keys. Additionally, SoCs incorporate specialized co-processors like graphics processing units, AI accelerators, Direct Memory Access (DMA) controllers, cryptographic accelerators, and communication modules. These co-processors play pivotal roles in executing computations involving sensitive data stored in memory components, such as cryptographic operations performed by crypto-accelerators, data transfers facilitated by DMA, and the training of machine learning models using AI accelerators.

Despite the implementation of various security mechanisms in modern SoCs, including fabric access control, control register locks, memory isolation, and secure debugging, there remain limitations to these security measures. 
Considering inherent constraints associated with post-silicon hardware patching and hardware fuzzing results in detecting hardware-level memory vulnerabilities in complex designs, making hardware fuzzing a high-potential candidate for SoC security verification.

\section{Conclusion}
In this paper, we have focused on hardware-related memory-safety issues and explored the efficacy of innovative techniques, specifically hardware fuzzing, in preemptively identifying pre-fabrication vulnerabilities. We have highlighted their strengths and weaknesses in detecting memory vulnerabilities by systematically reviewing existing hardware fuzzing techniques. The analysis indicates that hardware fuzzing has the potential to reveal specific memory vulnerabilities overlooked by other methods. This paper not only contributes to the understanding of hardware-related memory safety but also presents potential directions for the future development of hardware fuzzers to enhance their capabilities in memory safety verification for addressing the evolving landscape of hardware vulnerabilities and ensuring the overall security of computing systems.

\section{ACKNOWLEDGMENTS}
Our research work was partially funded by Intel's Scalable Assurance Program, Deutsche Forschungsgemeinschaft (DFG) – SFB 1119 – 236615297, the European Union under Horizon Europe Programme – Grant Agreement 101070537 – CrossCon, the European Research Council under the ERC Programme - Grant 101055025 - HYDRANOS, the US Office of Naval Research (ONR Award \#N00014-18-1-2058), and the Lockheed Martin Corporation. This work does not in any way constitute an Intel endorsement of a product or supplier. Any opinions, findings, conclusions, or recommendations expressed herein are those of the authors and do not necessarily reflect those of Intel, the European Union, the European Research Council, the US Government, or the Lockheed Martin Corporation.

\def\refname{REFERENCES}

\vspace*{-8pt}

\begin{IEEEbiography}{Mohamadreza~Rostami} is a Computer Science Ph.D. student at Technical University of Darmstadt at Darmstadt, Hessen, 64277, Germany. His research interests include Hardware Security, Hardware Fuzzing, and Microarchitectural Vulnerabilities. Rostami received his M.Sc. in Communication Security and Cryptography from the University of Tehran, Iran. Contact him at \url{mohamadreza.rostami@trust.tu-darmstadt.de}.
\end{IEEEbiography}

\begin{IEEEbiography}{Chen~Chen}{\,}is a Ph.D. student in Computer Engineering at Texas A\&M University at College Station, Texas, 77843, United States. His research interests include Hardware Security, Hardware Fuzzing, and Formal Verification. Chen received his M.S. in Electrical Engineering from the Wisconsin---Madison. He is a student member at ACM. Contact him at \url{chenc@tamu.edu}.
\end{IEEEbiography}

\begin{IEEEbiography}{Rahul~Kande}{\,}is a Ph.D. student in Computer Engineering at Texas A\&M University at College Station, Texas, 77843, United States. His research interests include Hardware Security and Computer Architecture, with a focus on developing Hardware Fuzzers to detect security vulnerabilities. Kande received his B.Tech degree in Electronics and Communication Engineering with a minor in Computer Science and Engineering from the Indian Institute of Technology Guwahati, India. Contact him at \url{rahulkande@tamu.edu}.

\end{IEEEbiography}

\begin{IEEEbiography}{Huimin Li}{\,} is a Ph.D. student in Cyber Security at Delft University of Technology, Delft, 2628XE, Netherlands. Her research interests include Hardware Security, Deep Learning, and RISC-V. Li received her master's degree in Micro-electromechanical System from Northwestern Polytechnical University, China. Contact her at \url{H.Li-7@tudelft.nl}.
\end{IEEEbiography}

\begin{IEEEbiography}{Jeyavijayan (JV)~Rajendran}{\,}is an Associate Professor in the Department of Electrical and Computer Engineering at the Texas A\&M University at College Station, Texas, 77843, United States. His research interests include Hardware Security and Computer Security. Prof. Rajendran obtained his Ph.D. degree from New York University in August 2015. His research has won the NSF CAREER award, ACM SIGDA Outstanding Young Faculty award, IEEE CEDA Ernest Kuh Early Career award, Intel Security Academic Leadership award, and Office of Naval Research Young Investigator award. Contact him at \url{jv.rajendran@tamu.edu}.
\end{IEEEbiography}

\begin{IEEEbiography}{Ahmad-Reza Sadeghi} is a full Professor of Computer Science at the Technical University of Darmstadt at Darmstadt, Hessen, 64277, Germany. His research interests include hardware security, software security, and AI security. Prof. Sadeghi received his Ph.D. in privacy-protecting cryptographic protocols and systems from the University of Saarland in Saarbrücken, Germany. Since 2012, he has been the director of several Intel collaborative centers. He has received prestigious awards, among others, ACM SIGSAC, Intel academic leadership award, European Research Council advanced grant, and German IT security. Contact him at \url{ahmad.sadeghi@trust.tu-darmstadt.de}.
\end{IEEEbiography}


\begin{thebibliography}{1}
\bibitem{dessouky2019hardfails}
G. Dessouky et al. ``HardFails: Insights into Software-Exploitable Hardware Bugs'' in {\it 28th USENIX Security Symposium}, 2019, pp. 213--230.

\bibitem{dessouky2021chunked}
G. Dessouky, A. Gruler, P. Mahmoody, A. Sadeghi, and E. Stapf. ``Chunked-cache: On-demand and scalable cache isolation for security architectures'' in {\it NDSS Symposium}, 2022.

\bibitem{frassetto2018imix}
T. Frassetto, P. Jauernig, C. Liebchen, and A. Sadeghi. ``IMIX: In-Process Memory Isolation EXtension'' in {\it 27th USENIX Security Symposium}, 2018, pp. 83--97.

\bibitem{bahmani2021cure}
R. Frassetto et al. ``CURE: A Security Architecture with CUstomizable and Resilient Enclaves'' in {\it 30th USENIX Security Symposium}, 2021, pp. 1073--1090.

\bibitem{2atson2015cheri}
R. N. M. Watson et al. ``CHERI: A Hybrid Capability-System Architecture for Scalable Software Compartmentalization'' in {\it IEEE Symposium on Security and Privacy}, 2015, pp. 20-37.



\bibitem{kande2022thehuzz}
R. Kande et al. ``TheHuzz: Instruction Fuzzing of Processors Using Golden-Reference Models for Finding Software-Exploitable Vulnerabilities'' in {\it 31st USENIX Security Symposium}, 2022, pp. 3219--3236.

\bibitem{chen2022hypfuzz}
C. Chen et al. ``HyPFuzz: Formal-Assisted Processor Fuzzing'' in {\it 32nd USENIX Security Symposium}, 2023.

\bibitem{chen2023psofuzz}
C. Chen, V. Gohil, R. Kande, A. Sadeghi, and J. Rajendran. ``PSOFuzz: Fuzzing Processors with Particle Swarm Optimization'' in {\it IEEE/ACM International Conference on Computer-Aided Design}, 2023.

\bibitem{gohi2024mabfuzz}
V. Gohil, R. Kande, C. Chen, A. Sadeghi, and J. Rajendran. ``MABFuzz: Multi-Armed Bandit Algorithms for Fuzzing Processors'' in {\it IEEE Design, Automation and Test in Europe Conference}, 2024.

\bibitem{mohamad2024chatfuzz}
M. Rostami, M. Chilese, S. Zeitouni, R. Kande, J. Rajendran, and A. Sadeghi. ``Beyond Random Inputs: A Novel ML-Based Hardware Fuzzing'' in {\it IEEE Design, Automation and Test in Europe Conference}, 2024.



\bibitem{rfuzz}
K. Laeufer, J. Koenig, D. Kim, J. Bachrach, and K. Sen. ``RFUZZ: Coverage-Directed Fuzz Testing of RTL on FPGAs'' in {IEEE/ACM International Conference on Computer-Aided Design}, 2018, pp. 1-8.

\bibitem{difuzzrtl}
J. Hur, S. Song, D. Kwon, E. Baek, J. Kim, B. Lee. ``Difuzzrtl: Differential Fuzz Testing to Find CPU Bugs" in {IEEE Symposium on Security and Privacy}, 2021, pp. 1286-1303.

\bibitem{morfuzz}
J. Xu, Y. Liu, S. He, H. Lin, Y. Zhou, and C. Wang. ``MorFuzz: Fuzzing Processor via Runtime Instruction Morphing Enhanced Synchronizable Co-simulation" in {32nd USENIX Security Symposium}, 2023, pp. 1307-1324.

\bibitem{cascade}
F. Solt, K. Ceesay-Seitz, and K. Razavi. ``Cascade: CPU Fuzzing via Intricate Program Generation" in {33rd USENIX Security Symposium}, 2024.


\bibitem{socfuzz}
M. Hossain, A. Vafaei, K. Z. Azar, F. Rahman, F. Farahmandi, and M. Tehranipoor. ``SoCFuzzer: SoC Vulnerability Detection using
Cost Function enabled Fuzz Testing" in {Design, Automation \& Test in Europe Conference}, 2023, pp. 1-6.




































































\end{thebibliography}
\end{document}